\documentclass[onecolumn,prd,aps,amsmath,showpacs,amssymb]{revtex4}

\usepackage[latin1]{inputenc}
\usepackage{graphicx}
\usepackage{subfigure}
\usepackage{color}
\usepackage{dcolumn}
\usepackage{bm}

\begin{document}

\title{Effective models of two-flavor QCD: finite $\mu$ and $m_q$-dependence}
\author{T.~K\"ah\"ar\"a}
\email{topi.kahara@jyu.fi}
\affiliation{Department of Physics, University of Jyv\"askyl\"a, P.O.Box 35, FIN-40014 Jyv\"askyl\"a, Finland \\
and 
Helsinki Institute of Physics, P.O.Box 64, FIN-00014 University of Helsinki, Finland}
\author{K.~Tuominen\footnote{On leave of absence from Department of Physics, University of Jyv\"askyl\"a}}
\email{kimmo.tuominen@phys.jyu.fi}
\affiliation{CP$^{ \bf 3}${-Origins}, 
Campusvej 55, DK-5230 Odense M, Denmark \\
and
Helsinki Institute of Physics, P.O.Box 64, FIN-00014 University of Helsinki, Finland}
\begin{abstract}
We study effective models of chiral fields and Polyakov loop expected to describe the dynamics responsible for the phase structure of two-flavor QCD at finite temperature and density. We consider chiral sector described either using linear sigma model or Nambu-Jona-Lasinio model and study the phase diagram and determine the location of the critical point as a function of the explicit chiral symmetry breaking (i.e. the bare quark mass $m_q$). We also discuss the possible emergence of the quarkyonic phase in this model.
\end{abstract}

\maketitle

\section{Introduction}

Hadronic matter undergoes a phase transition from hadronic matter into a partonic matter at high temperatures or densities. To predict the equation of state, study the existence of possible critical point(s) in the $(T,\mu)$-- phase diagram and the properties of the phase transitions presents a theoretical challenge on studies based on the fundamental theory of strong interactions, QCD. To obtain some insight into the QCD dynamics of quarks in the nonperturbative domains, models like the Nambu--Jona--Lasinio (NJL) model have been developed. These models are based on the chiral symmetry of light quarks.

Based on generic effective theory methods, identification of relevant degrees of freedom and symmetries, effective models able to account for the two most important features of QCD, deconfinement and chiral symmetry breaking, have been developed and studied \cite{Mocsy:2003qw, Sannino:2004ix}. These models include both chiral fields and the Polyakov loop as relevant degrees of freedom. The chiral symmetry is effectively represented by a NJL model or linear sigma model (LSM), while the $Z_3$ symmetry relevant for deconfinement in pure gauge theory is described by a mean field potential for Polyakov loop. The key role is played by the interactions coupling these two sectors. In the limit of light quarks the dynamics are driven by the chiral degrees of freedom and the decrease in the chiral condensate as the temperature is increased results in increase of the Polyakov loop which in turn results in deconfinement and explains the intertwining of these two seemingly unrelated features to a single phase transition \cite{Fukushima:2003fw, Ratti, Rossner:2007ik, Ciminale:2007ei,Costa:2008yh}. On the other hand, if the current quark masses are taken large, chiral symmetry broken explicitly, the dynamics is close to that of pure gauge theory, i.e. dominated by the Polyakov loop. Nevertheless, at deconfimenent the interactions now lead to decrease in the chiral condensate and the two transitions again coincide. In real QCD approximate chiral symmetry is typically expected to play the dominant role.

In an earlier work we have compared the models of two-flavor QCD where the chiral sector is represented either with LSM or NJL model \cite{Kahara:2008yg} at finite temperature and density. We found that while at zero chemical potential both cases give practically coincident results, the relative uncertainties increase when finite quark densities are considered. In particular the predictions for the location of the possible critical point in the $(T,\mu)$--plane differ widely \cite{Kahara:2008yg}. As a further direction to study and constrain these models we considered the dependence on explicit chiral symmetry breaking, i.e. the value of the quark mass $m_q$, in these models \cite{Kahara:2009sq}. In this brief report we complete this previous study by extending the analysis to finite density. We briefly comment also on the possible emergence of a quarkyonic phase introduced in recent literature \cite{McLerran:2007qj}.
	  
The paper is organized as follows: in section \ref{models} we briefly recall the basic definitions of the two models we consider and explain the model parameters which allow for arbitrary pion mass. In section \ref{results} we present our main results and in section \ref{checkout} our conclusions and outlook.	  
	  	  
\section{The models}
\label{models}

\subsection{Models at the physical point $m_\pi \approx 140$ MeV}

As explained in the introduction, in this work we continue our study \cite{Kahara:2009sq} of the quark (or pion) mass dependence of the QCD phase diagram. We consider the PNJL and PLSM models which consist of a chiral part, a lattice fitted Polyakov potential and a simple interaction between the two. The study of these models is done in the mean field approximation. For a detailed
description of the derivation see \cite{Kahara:2009sq}, here we will simply state the resulting grand potential
\begin{equation}
\Omega = U_{\rm{chiral}}+U_\ell+\Omega_{\bar{q}q}.
\label{Omega}
\end{equation}

The two models differ only in the chiral part which corresponds either to the NJL or LSM models. The linear sigma model (LSM) consist of the sigma meson and the pions with their mutual interactions, and interactions with quarks. The NJL model on the other hand describes only quarks with an effective four-fermion interaction. In both models, the explicit chiral symmetry breaking is taken into account. 

The deconfining phase transition is included in both models through the mean field potential
\begin{eqnarray}
U_\ell\equiv U(\ell,\ell^\ast,T) = T^4 \left(-\frac{b_2(T)}{2}|\ell|^2-\frac{b_3}{6}(\ell^3+
\ell^{\ast 3})+\frac{b_4}{4}(|\ell|^2)^2\right),
\label{polyakov_potential}
\end{eqnarray}
where
\begin{eqnarray}
b_2(T)=a_0+a_1\left(\frac{T_0}{T}\right)+a_2\left(\frac{T_0}{T}\right)^2+a_3\left(\frac{T_0}{T}\right)^3,
\end{eqnarray}
and the constants $a_i$,$b_i$ are fixed to reproduce pure gauge theory thermodynamics with 
phase transition at $T_0=270$ MeV; We adopt the values determined in \cite{Ratti}, and 
shown for completeness in table \ref{parametertable}. Here $\ell$ is the gauge invariant 
Polyakov loop in the fundamental representation. Instead of the polynomial form for the Polyakov loop potential  (\ref{polyakov_potential}), also other possibilities exist. An example is the one introduced in \cite{Roessner:2006xn} which has the advantage that $\ell$ is always confined to the values between zero and one. However as discussed in \cite{Fukushima:2008wg} these two possible forms for the potential do not differ significantly at temperatures below 300 MeV, the region we are interested in. Further improvement to the potential would be the inclusion of $\mu$--dependence in the $T_0$ parameter as is done for example in \cite{Schaefer:2007pw} and \cite{Abuki:2008nm}; we will discuss this briefly in Sec. \ref{resultsC}.

The chiral potentials in (\ref{Omega}) are
\begin{eqnarray}
U_{\rm{chiral}} &=& \frac{\lambda^2}{4}\left(\left(\frac{M}{g}\right)^2-v^2\right)^2 - \frac{HM}{g}, {\rm{\quad for\,\,LSM}} \\
U_{\rm{chiral}} &=& \frac{(m_q-M)^2}{2G}, {\rm{\quad for\,\,NJL}}. 
\end{eqnarray}
At the physical pion mass the parameters in the above equations are fixed by the physical vacuum properties. In the LSM model $H=f_\pi m_\pi^2$ and $v^2 = f_\pi^2 - m^2_\pi/\lambda^2$, where $f_\pi=93$ MeV and $m_\pi=138$ MeV. The coupling $\lambda^2 \approx 20$ is determined by the tree level mass $m_\sigma^2=2\lambda^2f_\pi^2+m_\pi^2$, which is set to be 600 MeV. In the NJL model we fix the bare quark mass to be $m_q=5.5$ MeV and the coupling $G=10.08 $ GeV$^{-2}$. The constituent masses $M$ are related to the $\bar{q}q$ and $\sigma$ expectation values throught the relations $M=m_q-G\langle\bar{q}q\rangle$ in NJL and $M = g\langle\sigma\rangle$ in LSM. In the latter case the coupling constant $g$ is fixed to 3.3 corresponding to the baryon mass $\sim 1$ GeV. 

The final term in (\ref{Omega}) includes the interaction between the chiral and Polyakov sectors and reads
\begin{equation}
\label{omegaqqbar1}
\Omega_{\bar{q}q} = 
-2N_fT\int\frac{d^3p}{(2\pi)^3}\left({\rm{Tr}}_c\ln\left[1+Le^{-(E-\mu)/T}\right]+{\rm{Tr}}_c\ln\left[1
+L^\dagger e^{-(E+\mu)/T}\right]\right),
\end{equation}
where the Polyakov loop matrix is $L = \exp[-g_s A_0/T]$. The above contribution is of the same basic form for both PLSM and PNJL models with $E = \sqrt{\vec{p}^{\,\,2} + M^2}$, where $M$ is the constituent mass of the model in question as defined above.
The trace over color remains and using the definition of the Polyakov loop $\ell=\langle{\rm{Tr}}_c(L)\rangle/N_c$ and taking the Polyakov loop matrix $L$ corresponding to a static background field $A_0$, one obtains
\begin{eqnarray}
\Omega_{\bar{q}q} &=& 
-2N_fT\int\frac{d^3p}{(2\pi)^3}\left(\ln\left[1+3(\ell+\ell^\ast e^{-(E-\mu)/T})e^{-(E-\mu)/T}+e^{-3(E-\mu)/T}\right]\right.\nonumber \\
&& \left.+\ln\left[1+3(\ell^\ast+\ell e^{-(E+\mu)/T})e^{-(E+\mu)/T}+e^{-3(E+\mu)/T}\right]\right).
\label{omegaqqbar2}
\end{eqnarray}
The interaction potential $\Omega_{\bar{q}q}$ includes also a vacuum term omitted from equations (\ref{omegaqqbar1}) and (\ref{omegaqqbar2}) 
\begin{equation}
-6N_f\int\frac{d^3p}{(2\pi)^3}E\theta(\Lambda^2-|\vec{p}|^2). 
\end{equation}
This term is neglected in the PLSM model, but included in the PNJL model where it is controlled by the cut-off $\Lambda$, which we set at 651 MeV. A summary of the parameters at the physical point is shown in table \ref{parametertable}.

The thermodynamics of the models are determined by solving the 
equations of motion for the order parameters,
\begin{eqnarray}
\frac{\partial\Omega}{\partial M}=0, ~~\frac{\partial\Omega}{\partial\ell}=0, 
~~\frac{\partial\Omega}{\partial\ell^\ast}=0,
\end{eqnarray}
and then the pressure is given by evaluating the potential on the minimum, $p=-\Omega(T,\mu)$. We have chosen the constituent mass $M$ as a basic variable since this most conveniently allows us to discuss both models simultaneously. It is also straightforward to write the results in terms of the condensates $\langle\sigma\rangle$ and $\langle\bar{q}q\rangle$ since these are linearly related to $M$ in each case. At finite chemical potential the mean field potential $\Omega$ is complex due to the Polyakov loops and minimizing such a potential is meaningless. A simple way to overcome this problem is to treat the Polyakov loop parameters
$\ell$ and $\ell^\ast$ as independent real variables, which will be a sufficient approximation for our current analysis. Inaccuracies of this treatment as well as improved methods have been discussed for example in \cite{Rossner:2007ik,Roessner:2006xn}.
 
\begin{table}[hbt]
\caption{The parameters used for the effective potential}
\begin{ruledtabular}
\begin{tabular}{llll}
 {\bf{LSM:}} & $f_\pi$ & $m_\pi$ & $m_\sigma$ \\
 & 93 MeV & 138 MeV & 600 MeV \\
 & $g$ &  $\lambda$ & $H$ \\
 & 3.3 & $\approx$ 4.44 & $\approx 1.77 \cdot 10^{-3}$ $\textrm{GeV}^3$ \\
\hline
 {\bf{NJL:}} & $m_q$ & $\Lambda$ & $G$ \\
 & 5.5 MeV & 651 MeV & 10.08 (GeV)$^{-2}$ \\
\hline
 {\bf{Polyakov:}} & $a_0$ & $a_1$ & $a_2$ \\
 & 6.75 & -1.95 & 2.625 \\
 & $a_3$ & $b_3$ & $b_4$ \\
 & -7.44 & 0.75 & 7.5 \\
\end{tabular}
\end{ruledtabular} 
\label{parametertable}
\end{table}

\subsection{Models away from the physical point}

To explore the region of parameter space in which the pion mass differs from its physical value, we need a consistent way of setting the model parameters in a such region. In the PNJL model this is quite easy since the bare quark mass is a direct input parameter that controls the amount of chiral symmetry breaking in the model and the pion mass is then calculated from the model. In the PLSM case the situation is not so
simple since we have four parameters, $f_\pi$, $m_\pi$, $m_\sigma$ and $g$, which are connected with each other. To make the comparison with the PNJL model easier we introduced in \cite{Kahara:2009sq} a parametrization based on lattice results \cite{Chiu:2003iw,Kunihiro:2003yj,Procura:2003ig} which relates the above parameters directly to the bare quark mass $m_q$. Here we will give a brief summary of this parametrization.

The lattice data connects the pion decay constant, pion mass and sigma mass through the equations
\begin{eqnarray}
m_\pi^2a^2 &=& (A_1(m_q a)^{\frac{1}{1+\delta}} + B(m_q a)^2) \\
\sqrt{2}f_\pi a  &=& 0.06672 + 0.221820\times (m_q a) - \sqrt{2}Ca \\
m_\sigma &=& \xi m_\pi^2 + D
\end{eqnarray}
obtained from \cite{Chiu:2003iw} and \cite{Kunihiro:2003yj}. The parameters in the above equations are shown in table \ref{parametertableII} and chosen to reproduce the physical vacuum values for
the PLSM model parameters i.e. $f_\pi = 93$ MeV, $m_\pi = 138$ MeV and $m_\sigma = 600$ MeV
when the quark mass $m_q$ is set to 5 MeV.

\begin{table}[hbt]
\caption{The lattice parameters}
\begin{ruledtabular}
\begin{tabular}{lllll}
$a$ 										& $A_1$   					& $\delta$ & $B$     					& $C$      \\
0.505306 (GeV)$^{-1}$		&	0.82725 					&	0.16413	 & 1.88687 					& 1.18 MeV \\
$D$ 										& $\xi$   					& $M_0$		 & $C_1$   					& $g_A$	  \\
565.15 MeV							& 1.83 (GeV)$^{-1}$ & 868 MeV  & 0.9 (GeV)$^{-1}$ & 1.267 
\end{tabular}
\end{ruledtabular} 
\label{parametertableII}
\end{table}

The PLSM model parameter $g$ is determined through the relation $gf_\pi = M_N/3$, where the nucleon mass $M_N$ is parametrized in the form
\begin{equation}
M_N = M_0 + 4C_1 m_\pi^2 - \frac{3g_A^2}{32\pi f_{\pi}^2}m_{\pi}^3.
\label{nuc_mass_par}
\end{equation}
This is a chiral perturbation theory fit truncated to ${\mathcal{O}}(m_\pi^3)$. In \cite{Procura:2003ig}
it has been shown that for a good description of nucleon mass one should keep terms up to and including ${\mathcal{O}}(m_\pi^4)$, but for simplicity we have chosen the truncated fit. Previously, in \cite{Kahara:2009sq}, we used the same nucleon mass formula truncated to ${\mathcal{O}}(m_\pi^2)$ but this exaggerated the strength of the coupling $g$ at larger $m_q$ to an extent that was found to have a large effect on the phase diagram; we chose to improve by adding the ${\mathcal{O}}(m_\pi^3)$-term. The parameters are chosen so that the PLSM model vacuum values are reproduced and the corresponding parameter values are shown in table \ref{parametertableII}. 

\begin{figure}[htb]
\centering
  \subfigure{
  \hskip-1.0truecm
  \includegraphics[width=8.5cm]{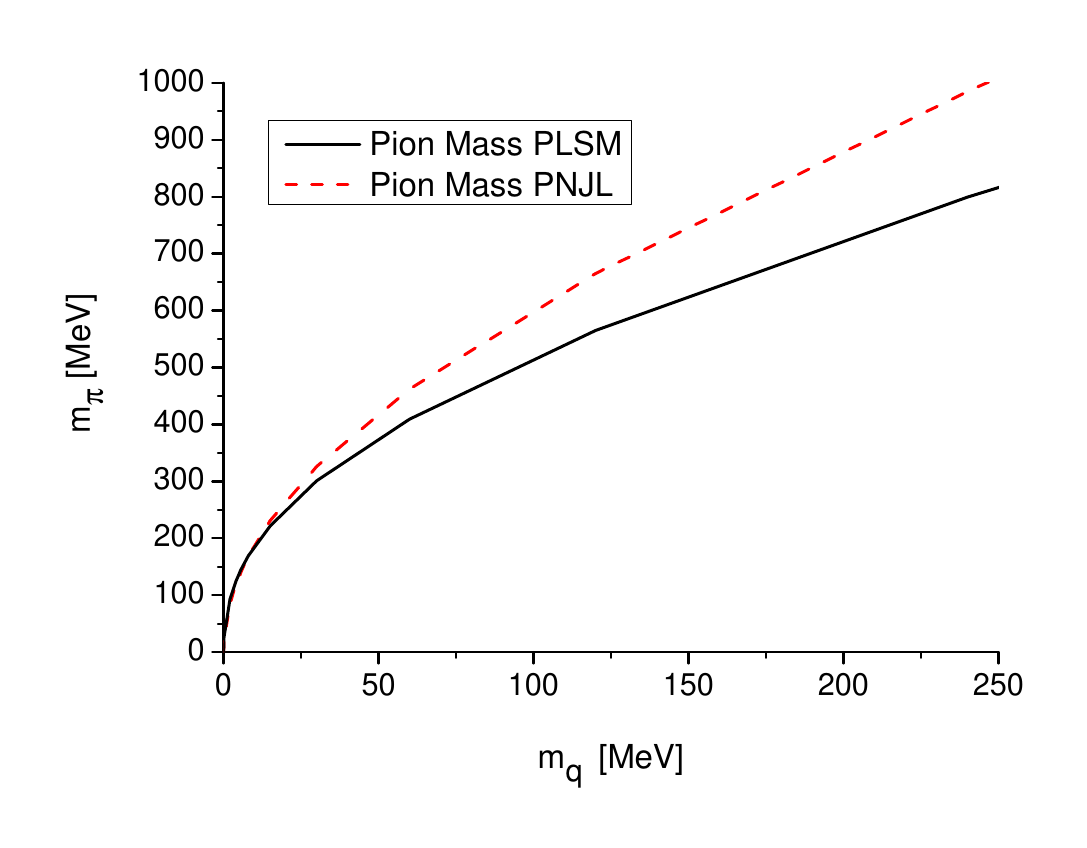}}
	\qquad
	\hskip-1.2truecm
	\subfigure{\includegraphics[width=8.5cm]{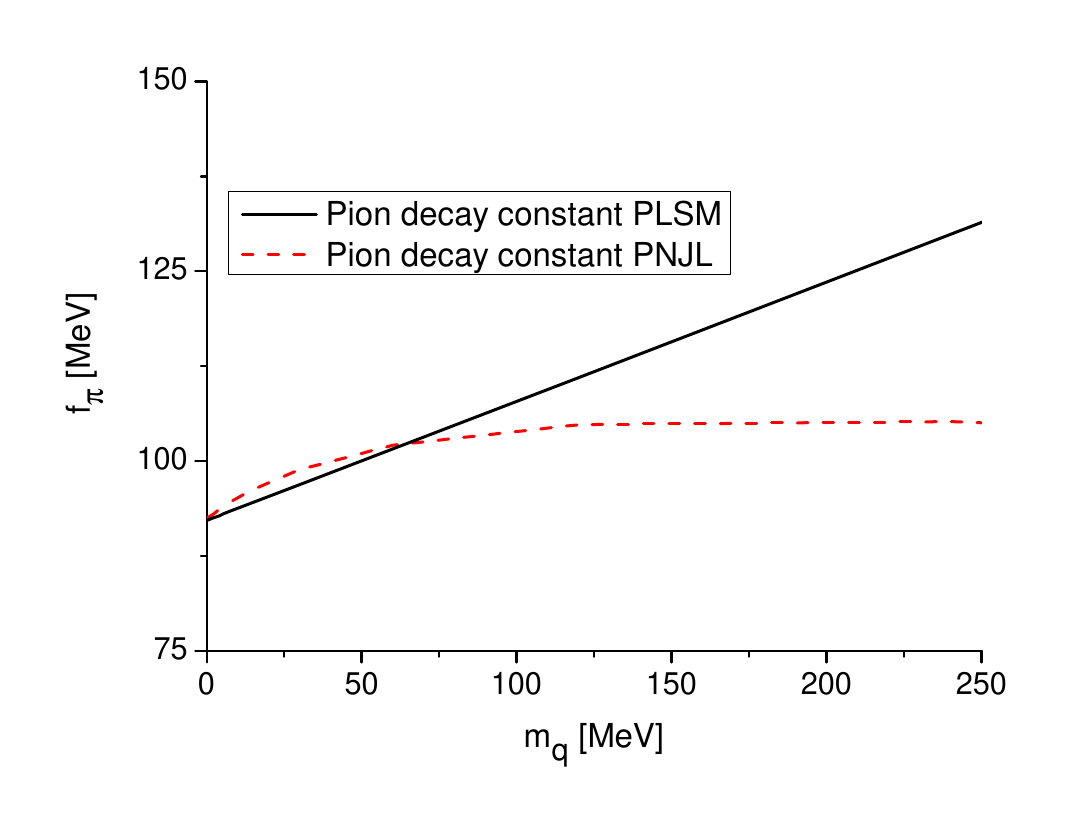}}
\vskip-0.4truecm
\caption{Pion mass $m_\pi$ and pion decay constants $f_\pi$ as functions of the bare quark mass $m_q$ in the PLSM and PNJL models.}
\label{mqvsmp}
\end{figure}

A comparison between the PNJL model and the PLSM model, now equipped with our lattice based parameter fit, is shown in Figure \ref{mqvsmp}. Both the pion masses and the pion decay constants agree very well between the two models at low quark masses. As the bare quark mass, i.e. the amount of explicit chiral symmetry breaking is increased, the models start to deviate. At $m_q = 250$ MeV, the largest quark mass shown in Figure \ref{mqvsmp}, the deviation is around 20 $\%$ for both the pion masses and decay constants and keeps increasing as one increases bare quark mass further. This growing deviation is a natural indication that the models, based on approximate chiral symmetry, start to fail as the explicit chiral symmetry breaking becomes large. It should be noted that the comparison between the PNJL and PLSM models includes no tuning of the coupling $G$ or the cut-off $\Lambda$ of the PNJL model. Actually we have checked numerically that the agreement between the PNJL and the lattice fitted PLSM pion mass curves in Figure \ref{mqvsmp} cannot be improved by altering the values of $G$ or $\Lambda$ from those shown in Table \ref{parametertable}.

In the following section we will study mainly the effect of explicit chiral symmetry breaking on the thermodynamics of the models. We focus in particular on the $(T, \mu)$-- phase diagram including possible critical points.

\begin{table}[hbt]
\caption{Bare quark masses with corresponding pion masses and critical points for both models. All values are in MeV.}
\begin{ruledtabular}
\begin{tabular}{llllll}
\multicolumn{3}{l}{\bf{PLSM}}   &  \multicolumn{3}{l}{\bf{PNJL}} \\
\hline
$m_q$ & $m_\pi $ & CP $(T,\mu)$ & $m_q$ & $m_\pi$ & CP $(T,\mu)$ \\
\hline
 0.1   & 26       & None         & 0.1   & 19      & (147, 270)   \\
  
 2     & 93       & None         & 2     & 85      & (111, 306)   \\
   
 5     & 138      & (196, 135)   & 5.5   & 140     & (\phantom{0}88, 329)   \\
   
 15    & 222      & (168, 235)   & 15    & 231     & (\phantom{0}59, 364)   \\
  
 50    & 377      & (120, 353)   & 50    & 421     & (\phantom{0}58, 435)   \\
 
 100   & 518      & (\phantom{0}30, 455)   & 100   & 603     & (\phantom{0}92, 496)   \\
  
 150   & 629      & None		     & 150   & 752     & (120, 536)   \\
 
\end{tabular}
\end{ruledtabular}
\label{mqmpitable}
\end{table}  

\section{Results}
\label{results}

By construction, in these models there are a priori two transitions: The chiral transition due to the (approximate) restoration of chiral symmetry and the deconfinement transition encoded into the Polyakov potential. The transitions can be studied through their respective order parameters, the constituent quark mass $M$ and the thermal average of the Polyakov field, $\ell$. To both transitions one can assign their own critical temperatures. The definition of the critical temperature, however, is vague especially in the regions where the transition is a crossover and the order parameter shifts continuously. Since this is the case over large portion of the $(T,\mu)$--plane, in this work we primarily define the transition temperature as the temperature at which the temperature derivative of the order parameter has a maximum. Even this definition has some problems, since in some cases the derivative has several local maxima indicating rapid changes at several different temperatures. The critical temperature is identified with the maximum at which the change in the absolute value of the corresponding order parameter is largest. Alternatively one could use the susceptibilities to define the critical temperature. 

\subsection{The chiral phase diagram and the critical point}
\label{resultsA}

\begin{figure}[htb]
\centering
  \subfigure{
  \hskip-1.0truecm
  \includegraphics[width=8.5cm]{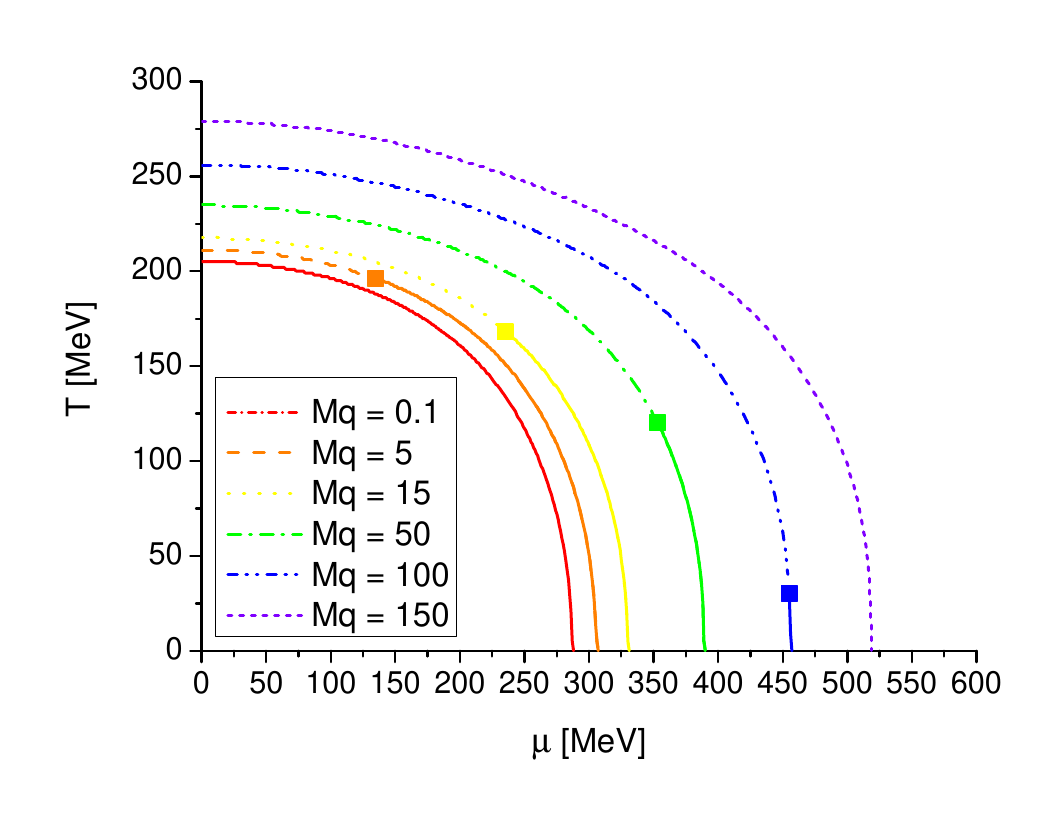}}
	\qquad
	\hskip-1.2truecm
	\subfigure{\includegraphics[width=8.5cm]{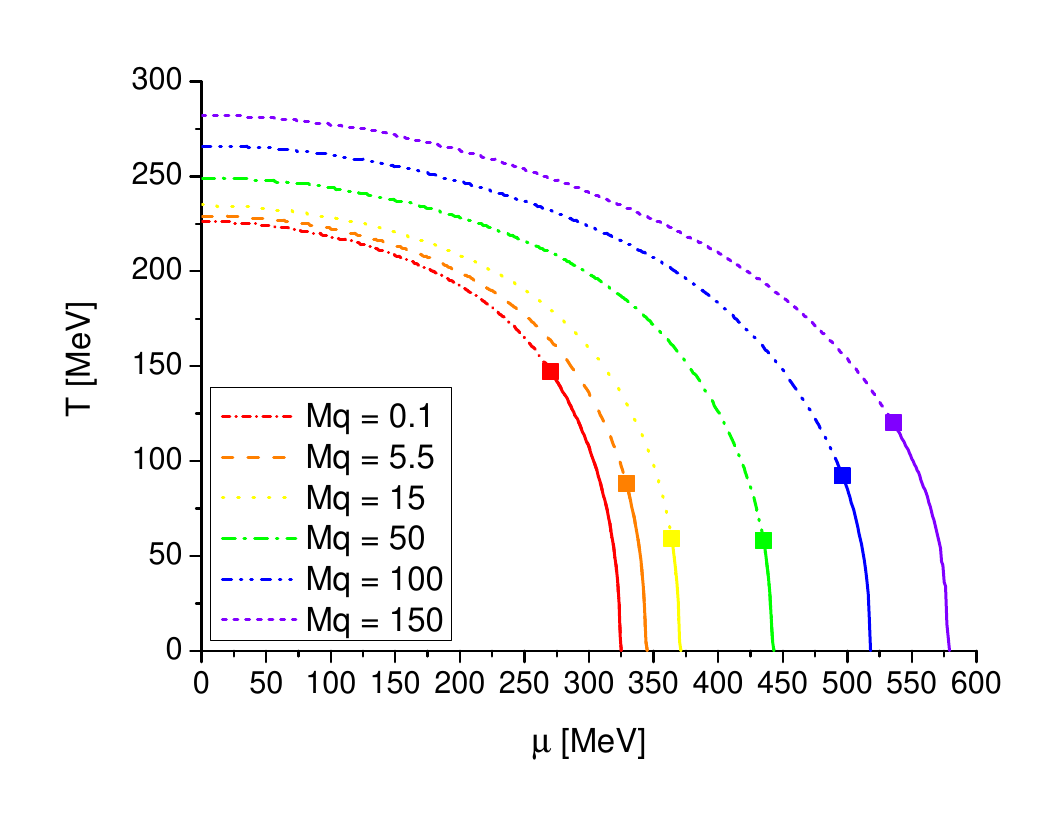}}
\vskip-0.4truecm
\caption{The chiral phase diagrams of the models for several quark masses. The solid curves indicate a first order (discontinuous)
transition with the critical endpoints marked by squares. Left: PLSM Right: PNJL}
\label{CMPDG}
\end{figure}

The chiral transition can be determined by finding the temperature corresponding to the fastest change in the constituent quark mass $M$ at fixed chemical potential $\mu$ (or vice versa). This transition temperature corresponds in most cases to the temperature at which the constituent mass drops below $50 \%$ of its vacuum value, only at large $m_q$ and $\mu$ does the fastest change occur at a different temperature than the one where the decrease in the absolute value of the constituent mass takes place. Figure \ref{CMPDG} shows the chiral transition lines in the $(T,\mu)$--plane for different quark masses for both models; also the critical points are shown. The critical points indicate the points where a line of first order (discontinuous) transitions ends and turns into a crossover (continuous). 

As seen in Figure \ref{CMPDG} the qualitative features of the phase diagrams in the two models are very similar: As the quark mass rises, the area under the transition line expands and the critical point moves towards larger $\mu$. The quantitative difference in the transition temperature between the models is below $15 \%$ for the shown quark masses. However the critical points appear to be located quite differently in the models as we already noted for the physical value of $m_q$ in \cite{Kahara:2008yg}. In the PLSM the transition for the lowest quark masses shown in Figure \ref{CMPDG} is first order all the way so there is no critical point in the $(T,\mu)$--plane, the same holds for the largest quark mass but now the transition is a crossover throughout the plane. Therefore, in the PLSM model the critical point is present only at a finite $m_q$ interval and outside this interval the transition is either entirely crossover or entirely of first order. In \cite{Schaefer:2008hk} a qualitatively similar result has been obtained for a three flavor linear sigma model. The fact that the transition in PLSM is of first order over the entire $(T,\mu)$--plane in chiral limit is due to neglect of the fermion vacuum energy 
\cite{Skokov:2010sf}.

In the PNJL case, where the fermion vacuum contribution is included, the critical points persist even for the smallest quark masses shown and the transitions at zero chemical potential remain crossovers. Also the critical temperature at the critical point starts to rise again at larger quark masses and does not disappear as in the PLSM case. A similar effect has been observed in \cite{Roessner:2006xn}, where a saturation of the critical point temperature was mentioned and attributed to a diquark dominated phase. Since our work does not include diquark degrees of freedom, we conclude that the behaviour of the critical point at large quark masses is a more generic feature of the PNJL model.
   
It has been suggested in \cite{Bowman:2008kc} that there might be, especially at small quark masses, multiple critical points in the $(T,\mu)$--plane.  We, however, found no evidence in either model to suggest that this is the case. In our previous work \cite{Kahara:2009sq} we noted that at large quark masses the PLSM transition was first order at $\mu = 0$ giving some credence to the idea of multiple critical points at large quark masses. 
As mentioned in the previous paragraph and evident from Figure \ref{CMPDG}, this is not the case in our present work and the first order transition at $\mu = 0$ observed in \cite{Kahara:2009sq} was caused by the overestimation of the nucleon mass $M_N$ at large pion masses and the resulting overestimation of the coupling $g$. Now, with the more precise formula (\ref{nuc_mass_par}) for the nucleon mass, the transition is a crossover for all quark masses above the physical mass $m_q = 5$ MeV. 

\subsection{The deconfinement transition}
\label{resultsB}

The deconfinement of the system is quantified by the Polyakov loop order parameter $\ell$ and its conjugate $\ell^\ast$, which we treat as independent real variables. As with the chiral transition the transition temperatures could be determined locating the maxima of the temperature derivatives. However, as noted in our previous works \cite{Kahara:2008yg,Kahara:2009sq}, there is in some cases a double peak structure in the derivatives with one peak coinciding with the chiral transition and caused by the interaction between the chiral and Polyakov sectors of the models. The second peak is a softer one and is related to the the transition present in the parametrization of the Polyakov loop mean field potential. 
Generally, the softer peaks occur at the values $\ell = 1/2$ and $\ell^\ast = 1/2$, so they could be considered as measures of the system becoming deconfined. This is a reasonable statement since confinement in the models is due to numerical suppression of quarks states through the order parameters $\ell$ and $\ell^\ast$. This means that alternatively the deconfinement could be considered to occur when the suppressing order parameters reach large enough value in order not to provide suppression anymore. Since the Polyakov loop order parameters $\ell$ and $\ell^\ast$ obtain, with our choice of potential, values roughly from 0 to 1, the values $\ell = 1/2$ and $\ell^\ast = 1/2$ are a natural choice to be the indicator of when the system turns from a mostly confined state to a mostly deconfined state, bearing in mind that the transition is a crossover.
To further simplify the analysis and readability of the figures we will, since  $\ell$ and $\ell^\ast$ do not coincide at finite $\mu$, define single deconfinement transition temperature as the average of the transition temperatures determined by $\ell=1/2$ and $\ell^\ast=1/2$.

\begin{figure}[htb]
\centering
  \subfigure{
  \hskip-1.0truecm
  \includegraphics[width=8.5cm]{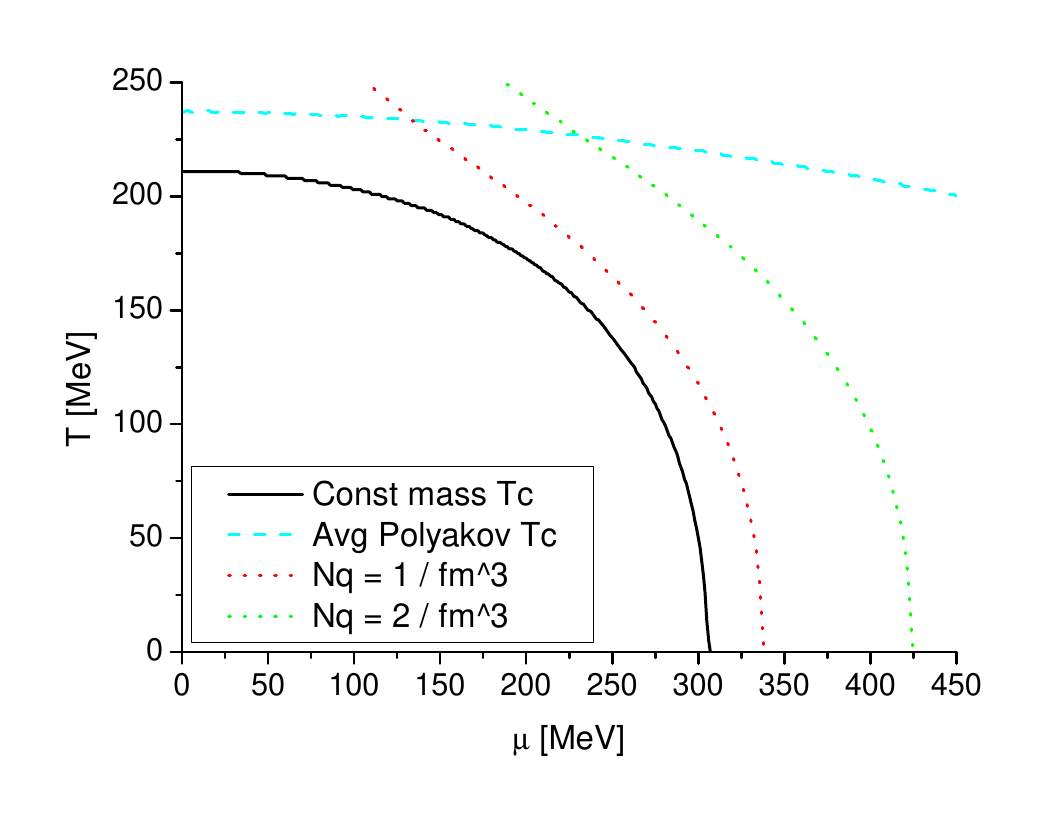}}
	\qquad
	\hskip-1.2truecm
	\subfigure{\includegraphics[width=8.5cm]{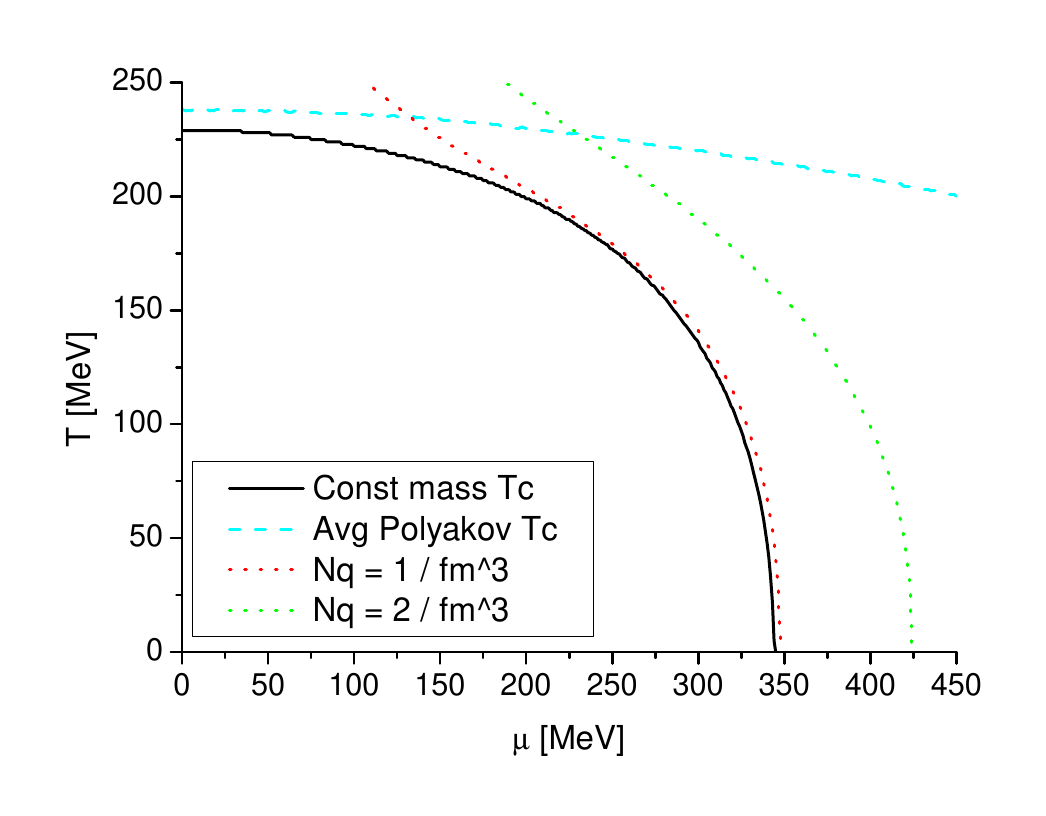}}
\vskip-0.4truecm
\caption{The phase diagrams of the models at the physical point $m_q \approx 5$ MeV with no explicit $\mu$--dependence in the
Polyakov potential. The solid line is the chiral transition and the dashed line the deconfinement transition.
The dotted lines correspond to quark number densities of 1 fm$^{-3}$ and 2 fm$^{-3}$.
Left: PLSM Right: PNJL.}
\label{QNDG}
\end{figure}

\begin{figure}[htb]
\centering
  \subfigure{
  \hskip-1.0truecm
  \includegraphics[width=8.5cm]{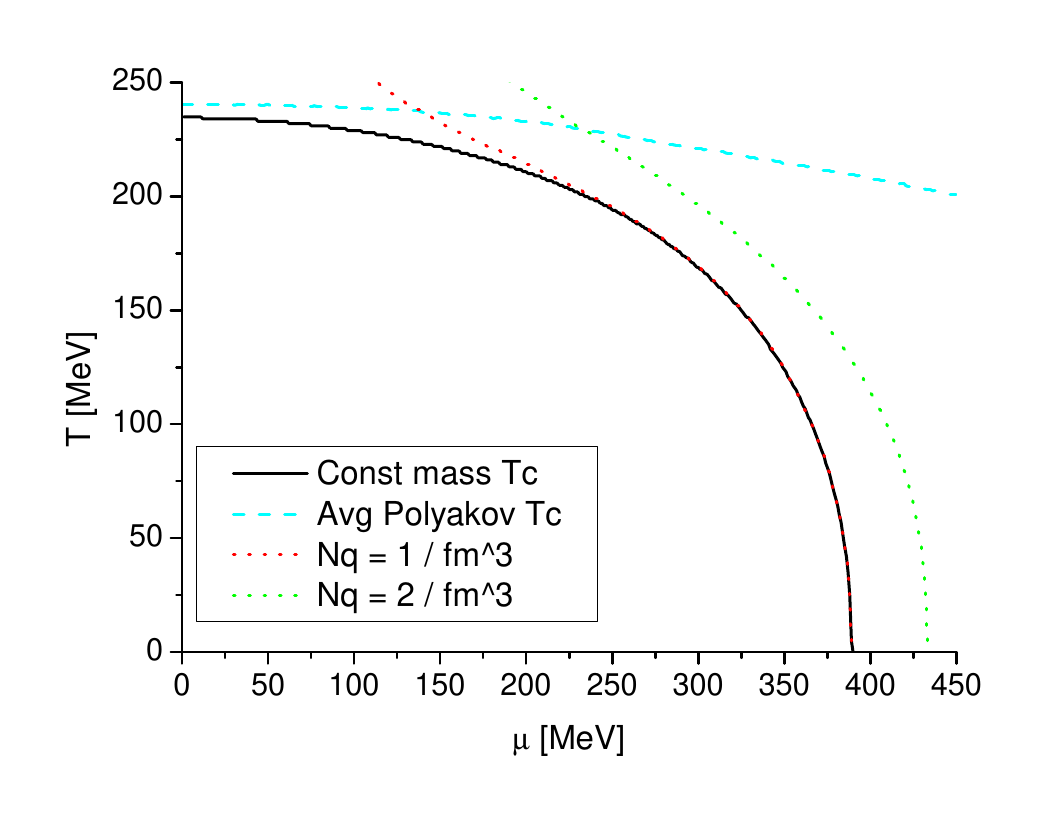}}
	\qquad
	\hskip-1.2truecm
	\subfigure{\includegraphics[width=8.5cm]{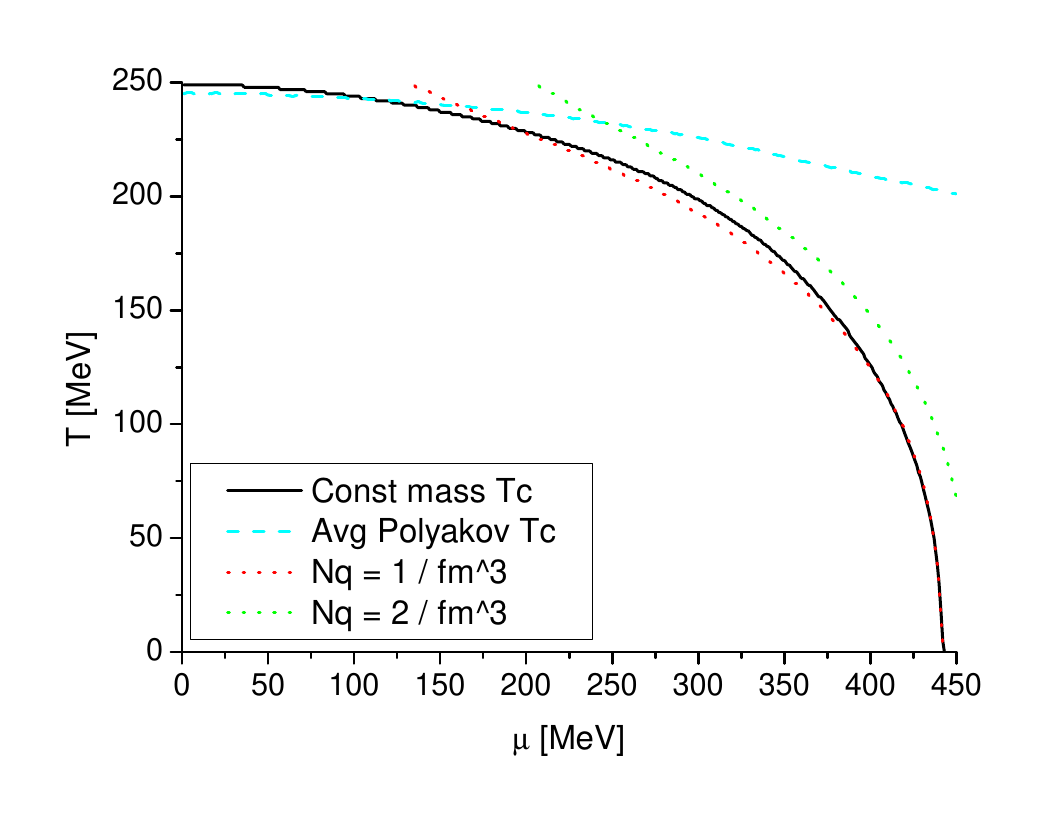}}
\vskip-0.4truecm
\caption{The phase diagrams of the models at $m_q = 50$ MeV with no explicit $\mu$--dependence in the
Polyakov potential. The solid line is the chiral transition and the dashed line the deconfinement transition.
The dotted lines correspond to quark number densities of 1 fm$^{-3}$ and 2 fm$^{-3}.$
Left: PLSM Right: PNJL.}
\label{QNDG50}
\end{figure}

In Figures \ref{QNDG} and \ref{QNDG50} the averaged deconfinement transition lines are shown for the physical quark mass $m_q \approx 5$ MeV and a larger quark mass $m_q = 50$ MeV along with the corresponding chiral transitions.
The first observation is that, defined this way, the deconfinement transition is independent of the chiral model
and also of the amount of explicit chiral symmetry breaking i.e. the quark mass. This means that tuning the Polyakov potential, so that one obtains a coincidence of the deconfining and chiral transitions for one chiral model and a specific quark mass, will not give the same outcome in other cases. Furthermore, the deconfinement transition line, $T_{c,\rm{dec}}(\mu)$, depends only very weakly on $\mu$. This means that deconfinement and chiral restoration can be made to coincide only at $\mu=0$. However, on the basis of symmetries one does not expect deconfinement and chiral restoration to become independent at finite $\mu$. 
Hence one is led to study possible $\mu$-dependence in the Polyakov loop potential.

To illustrate the effects of $\mu$--dependence of the Polyakov potential on the deconfinement transition, we adopt the formulation from \cite{Schaefer:2007pw} with the following modifications: In \cite{Schaefer:2007pw} the $\mu$--dependence of the Polyakov
potential was through the critical temperature $T_0$, which could be described by the following parametrization
\begin{equation}
T_0(\mu) = T_\tau e^{-1/(\alpha_0 b(\mu))},
\end{equation}   
with the coefficient $b(\mu)$ depending on the number of colors, massless flavors and the chemical potential,
\begin{equation}
b(\mu) = \frac{11N_c - 2N_f}{6\pi} - \frac{16N_f}{\pi} \frac{\mu^2}{T_\tau^2}.
\end{equation}
The parameters $\alpha_0 = 0.304$ and $T_\tau = 1.770$ GeV were fixed to reproduce the $N_f = 0$ lattice result $T_0(\mu = 0) = 270$ MeV.
For two massless flavors this would mean $T_0(\mu = 0) = 208$ MeV. However, since we want to make a comparison between the the
$\mu$--dependent and the $\mu$--independent cases, we keep the $\mu = 0$ point as a reference point for the two cases. Hence we use
\begin{equation}
b(\mu) = \frac{11N_c}{6\pi} - \frac{16N_f}{\pi} \frac{\mu^2}{T_\tau^2}.
\label{our_b}
\end{equation}
This modification does not alter the fixing of the
parameters $\alpha_0$ and $T_\tau$, which we fix to their above mentioned values. Our parametrization implies $T_0(\mu = 0) = 270$ MeV for any number of flavors. However, we are not interested in $N_f$-dependence since will exclusively consider the case $N_f=2$ and we want to only use the $\mu$-dependent Polyakov loop potential to illustrate the uncertainties which may arise in our consideration of $(T,\mu)$-- phase diagrams. The $N_f$-dependence in (\ref{our_b}) also neglects the effect of quark masses, which will suppresses $T_0$ as discussed in \cite{Schaefer:2007pw}. This effect is not large for two light flavors, but grows more significant when considering larger quark masses. We stress that our intention here is simply to illustrate the effect of $\mu$-dependent Polyakov potential on the phase diagram and for that purpose the simple parametrization we have chosen is sufficient.

\begin{figure}[htb]
\centering
  \subfigure{
  \hskip-1.0truecm
  \includegraphics[width=8.5cm]{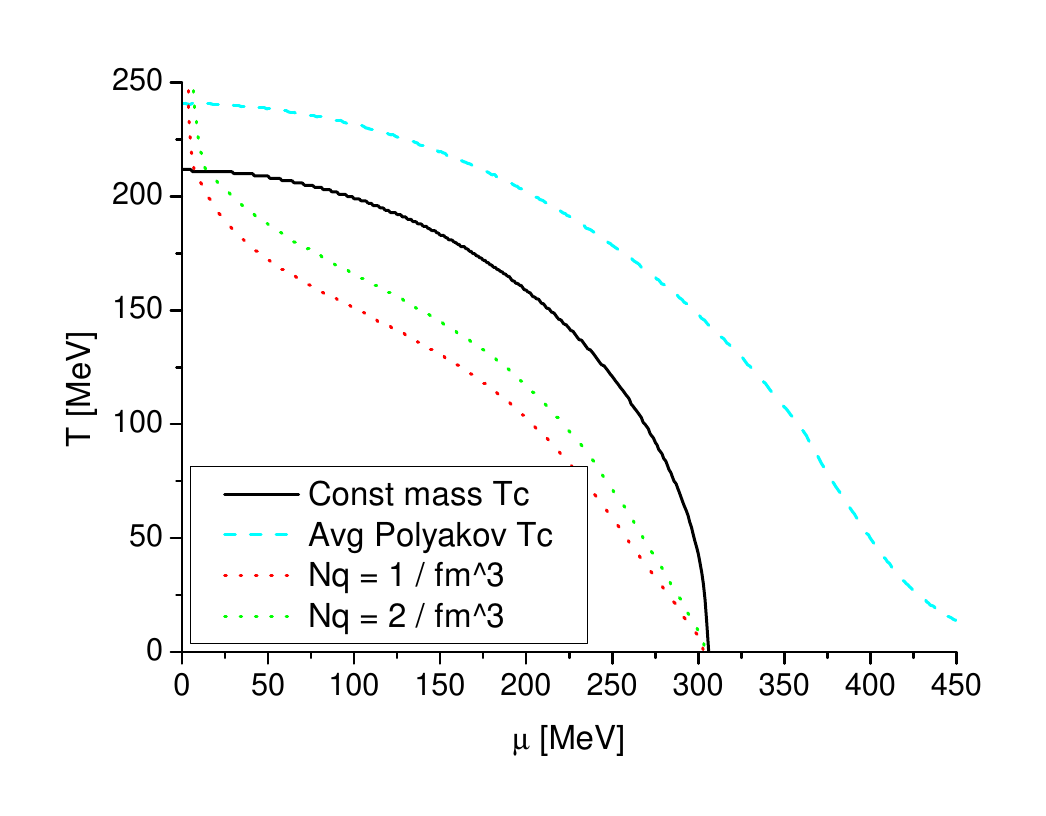}}
	\qquad
	\hskip-1.2truecm
	\subfigure{\includegraphics[width=8.5cm]{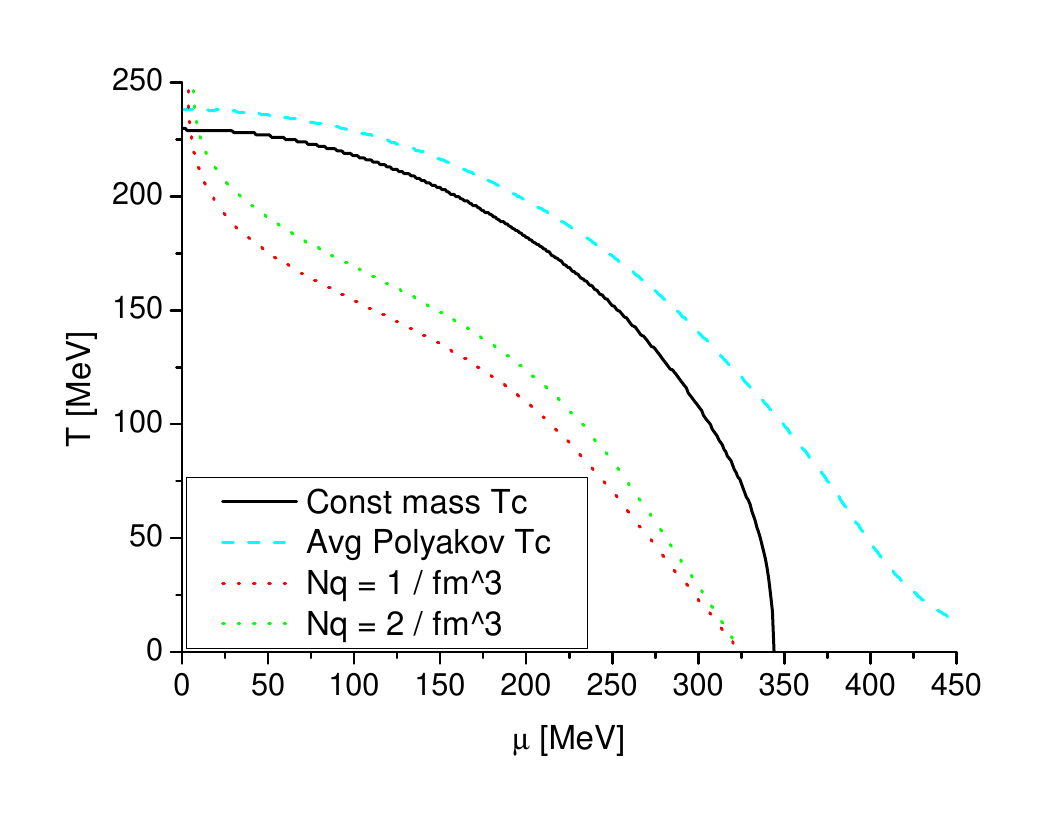}}
\vskip-0.4truecm
\caption{The phase diagrams of the models at the physical point $m_q \approx 5$ MeV with a $\mu$--dependent
Polyakov potential. The solid line is the chiral transition and the dashed line
the deconfinement transition. The dotted lines correspond to quark number densities of 1 fm$^{-3}$ and 2 fm$^{-3}$.
Left: PLSM Right: PNJL (Color online)}
\label{QNDGMD}
\end{figure}

\begin{figure}[htb]
\centering
  \subfigure{
  \hskip-1.0truecm
  \includegraphics[width=8.5cm]{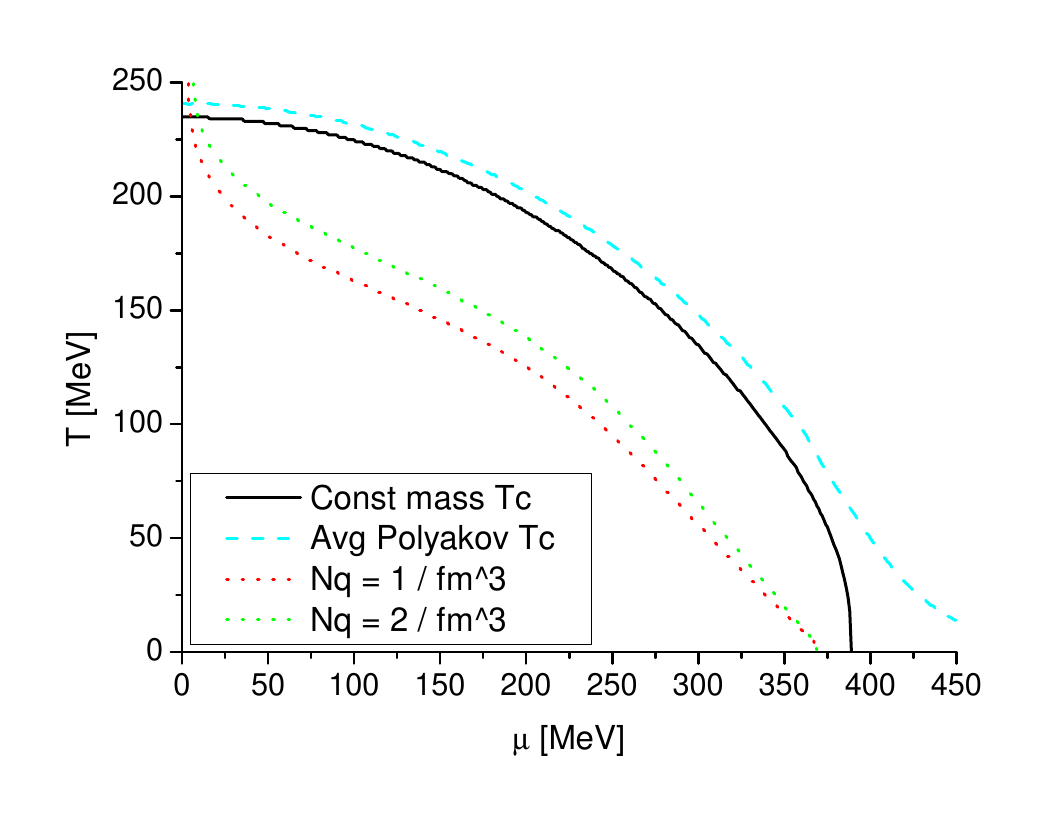}}
	\qquad
	\hskip-1.2truecm
	\subfigure{\includegraphics[width=8.5cm]{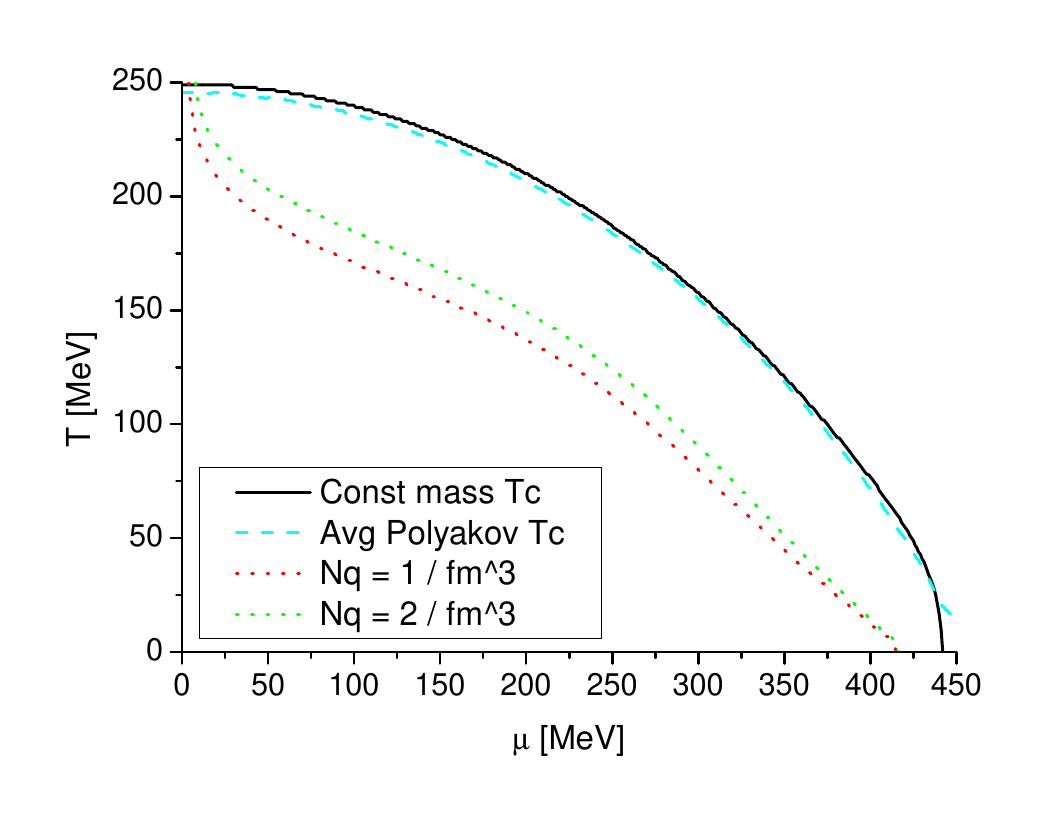}}
\vskip-0.4truecm
\caption{The phase diagrams of the models at $m_q = 50$ MeV with a $\mu$--dependent
Polyakov potential. The solid line is the chiral transition and the dashed line
the deconfinement transition. The dotted lines correspond to quark number densities of 1 fm$^{-3}$ and 2 fm$^{-3}$.
Left: PLSM Right: PNJL (Color online)}
\label{QNDG50MD}
\end{figure}

The phase diagrams obtained using the $\mu$--dependent Polyakov potential are shown in  Figures \ref{QNDGMD} and \ref{QNDG50MD}. The deconfinement transition now appears very different: The main new feature is that the deconfinement temperature follows the chiral restoration critical temperature more closely. Of course one may argue that this feature is put in by hand into the models of this type, but one the other hand the deconfinement phase transition is put in by hand already into the $\mu=0$ Polyakov potential. Lattice determination of the coincidence of the chiral symmetry restoration and deconfinement at finite $\mu$ would provide strong motivation to use $\mu$-dependent Polyakov potential in these effective modes. However, on the basis of the symmetries of the underlying gauge dynamics, one would indeed expect deconfinement and chiral symmetry restoration to coincide also at finite $\mu$.

As a final application we will discuss the possibility of describing the quarkonic phase with these models.

\subsection{Quarkyonic matter?}
\label{resultsC}
Recently there has been some interest towards a novel form of matter conjectured to exist in the QCD phase diagram \cite{McLerran:2007qj}. To briefly recall, the conjecture is based on considerations at 't Hooft large $N_c$ limit and rests mainly on the following two features: First, since the free energy scales with the number of degrees of freedom, one has hierarchical contributions from the quarks and gluons, of the order of $N_c$ and $N_c^2$, respectively. Second, while the mesons and glueballs become free in the large $N_c$ limit, their cubic and quartic interactions vanishing as $1/\sqrt{N_c}$ and $1/N_c$ respectively, the baryons remain strongly coupled. Both these features are, however, valid in the large $N_c$ limit while for QCD $N_c=3$. Furthermore, as we will now briefly discuss, the large $N_c$ limit is not unique \cite{Corrigan:1979xf} .

In the 't Hooft limit of large $N_c$ the fermions are taken to transform according to the fundamental representation. This leads to the well known features at large $N_c$: Planar diagrams dominate, and among these diagrams quark loops are suppressed relative to gluonic ones. As the dimension of fermion representation is $N_c$ while the one for gluons is $N_c^2-1\simeq N_c$, different orders of magnitude for the free energies emerge as $N_c$ is taken large. 

However, consider the following redefinition of the quark fields
\begin{equation}
Q^a = \epsilon^{abc} Q^{bc},\,\,\,Q^{bc}=-Q^{cb},
\end{equation}
i.e. consider quarks to transform in the two index anti-symmetric representation, which for SU(3) conceptually corresponds to renaming antiquarks as quarks. While this changes nothing in the dynamics for $N_c=3$, the large $N_c$ limit is entirely different. This is due to the fact that quarks in the anti-symmetric representation are counted similarly to gluons and hence they do not decouple at large $N_c$. The free energies of quarks and gluons are both of the order of $N_c^2$. The different behavior between mesons and baryons is similar to 't Hooft limit.
 
These different large $N_c$ limits emphasize different phenomenological features, and t is difficult to argue that either would be more realistic; more probably both are somewhat idealized and equally far from real three color QCD. And these two do not even exhaust the possible large $N_c$ limits. For example, one can also consider the possibility of having a hybrid large $N_c$ limit where out of three flavors one quark flavor transforms in the two-index antisymmetric representation while the other two are taken to transform in the fundamental representation. In this case it is possible to construct color singlet states consisting of three quarks at any $N_c$. These baryons are very different in comparison to baryons of the other two large $N_c$ limits considered above: their masses do not grow with $N_c$ and their Regge slopes coincide with the Regge slopes of mesons.

We do not embark on a thorough analysis of the phenomenology associated with these large $N_c$ limits; see e.g. \cite{Cherman:2009fh}. From the above discussion we simply remark that the extrapolations from large $N_c$ limits to real three color QCD should be taken with a grain of salt. A lattice study considering the quarkyonic phase in two-color QCD has recently appeared \cite{Hands:2010gd}.

It has been suggested that that the quarkyonic phase can be characterized by a non-vanishing baryon density while the system is still in a confined phase \cite{McLerran:2007qj}. In our models we have access to the quark number density
\begin{equation}
n_q = \frac{\partial \Omega}{\partial \mu}
\end{equation}
which we can use as a measure of the baryon density. 

If we first consider the case of a $\mu$--independent Polyakov potential, illustrated in Figures \ref{QNDG} and \ref{QNDG50}, there is a sharp rise in the quark number density simultaneously with the chiral transition. As with the deconfinement transition, the absolute magnitude of this rise is not great, it is around 0.6 fm$^{-3}$, but the relative increase is about a factor of four. So it could be argued that at the point of the chiral transition also the number density changes from a nearly zero value to a non-zero one. Another way of characterizing this 'quarkyonic transition' is to assign a threshold value for $n_q$ which separates the phases. This, however, is more arbitrary since the threshold value of $n_q$ one could choose is not in any way unique. In Figures \ref{QNDG} and \ref{QNDG50} we have plotted the chiral transition line, which corresponds to the rise in quark number density, along with the deconfinement transition line, obtained as explained in the previous 
section, and two curves which correspond to values of 1 and 2 fm$^{-3}$ of $n_q$. If one naively expects a baryon to have quark density around 3 fm$^{-3}$ then one should have nonzero net baryon density not later than when the average quark density hits 3 fm$^{-3}$, most likely even sooner. Seeing also that the deconfinement line lies at large temperatures, there is a substantial window for the quarkyonic matter to exist realizing the picture envisioned in \cite{McLerran:2007qj}.

However, this picture changes considerably if the Polyakov loop potential depends explicitly on $\mu$. 
This case is shown in 
Figures \ref{QNDGMD} and \ref{QNDG50MD}. First of all the quantitative behaviour of the quark number density changes quite drastically: Now
 $n_q=1$ and $n_q=2$ fm$^{-3}$ lines are well inside the chirally broken phase, but also the rise in quark number density associated with the chiral transition becomes significantly larger, in particular at large $\mu$. 
But more importantly, the
change in the deconfining transition, which now follows the chiral transition more closely, significantly decreases the area where possible quarkyonic matter
could reside. Here one should remember that when implementing the $\mu$--dependence to the Polyakov potential, we neglected effects from number of flavors and quark masses, which would bring the deconfinement lines down even faster than in Figures \ref{QNDGMD} and \ref{QNDG50MD} and thus practically closing the window for the existence of quarkyonic matter. 

\section{Conclusions}
\label{checkout}

We have considered the $(T,\mu)$-- phase diagram of two-flavor QCD in effective models which take into account both chiral degrees of freedom relevant for the restoration of the chiral symmetry and Polyakov loop relevant for deconfienement. Earlier these studies have been performed by constraining these models to reproduce the physical vacuum, in the two flavor case essentially determined by the bare quark mass $m_q=m_u=m_d$ (or alternatively by the pion mass $m_\pi$). We have relaxed this assumption and treated $m_q$ as a free parameter of the model in order to study how the explicit chiral breaking manifests in the thermodynamics.

We considered, side by side, two different models PNJL and PLSM which differ by the choice of the effective realization for the chiral sector. In earlier studies these two models have been shown to lead to qualitatively similar results for the thermodynamics both at finite temperature and density; the main quantitative difference has been shown to be in the location of the critical point in the $(T,\mu)$-- phase diagram. In this paper we have shown that, as a function of $m_q$, a qualitative difference arises: while in PNJL model the critical point exists for any $m_q$, in PLSM model the critical point exists only for a range of values of $m_q$. In both models we find, for any $m_q$, at most one critical point in contrast to the results in \cite{Bowman:2008kc} where multiple critical points were observed for LSM model at non-physical values of the pion mass.

Finally, we applied these effective models to consider the quarkyonic phase in two flavor QCD. The theoretical motivations on the existence of this novel phase are based on large $N_c$-limit of QCD and sensitive to {\em{which}} large $N_c$ limit is considered model studies are required. Furthermore, the existence of this novel state of matter was shown to depend sensitively on the parametrization of the Polyakov loop potential at finite $\mu$. In particular, for $\mu$--independent Polyakov loop potential there appears a wide window in the $(T,\mu)$-- phase diagram for quarkyonic matter to exist while if $\mu$ dependence in the Polyakov loop is introduced to obtain coincidence of chiral symmetry restoration and deconfinement similarly as at $\mu=0$, the window for quarkyonic matter practically closes.   

\acknowledgments
The financial support for T.K. from the V\"ais\"al\"a foundation is gratefully acknowledged.

\end{document}